# Design and analysis of electrothermal metasurfaces

Xiu Liu[†], Zhuo Li[†], Zexiao Wang[†], Hyeong Seok Yun, Sheng Shen[*]

Department of Mechanical Engineering, Carnegie Mellon University (Pittsburgh PA 15213, USA)

**Abstract** Electrothermal metasurfaces have attracted extensive attention due to their ability to dynamically control thermal infrared radiation. Although previous studies were mainly focused on the metasurfaces with infinite unit cells, the finite-size effect can be a critical design factor for developing thermal metasurfaces with fast response and broad temperature uniformity in practice. Here, we study the thermal metasurfaces consisting of gold nanorods with a finite array size, which, with only several periods, can achieve a resonance close to that of the infinite case. More importantly, such a small footprint due to the finite array size results in the response time down to a nanosecond level. Furthermore, the number of the unit cells in the direction perpendicular to the axis of the nanorods is found to be insensitive to the resonance and response time, thus providing a tunability in aspect ratio that can boost the temperature uniformity in the sub-Kelvin level.

**Keywords** modulated thermal infrared radiation, metasurface, nanosecond response time, sub-Kelvin temperature uniformity, finite size, aspect ratio

## 1 Introduction

Metasurfaces, consisting of an array of subwavelength scatters such as metallic nanorods[1–3], graphene nanoribbons[4,5], and dielectric gratings[6], have emerged as a promising platform to actively control thermal infrared radiation[7–12]. The coupling among these thermal scatters or infrared antennas provides rich degrees of freedom for the control of thermal emission, including its spectrum, directionality, and polarization. The resulting metasurface thermal emitters have become excellent alternatives to semiconductor based infrared light sources (e.g., infrared lasers or LEDs) in the applications of lighting[13–15], sensing[16], imaging[17], and energy harvesting[18,19], because of their high tunability and low-cost.

However, there still exist substantial challenges for the dynamic control of thermal infrared emission using active thermal metasurfaces, such as slow temporal response[13,20–22], poor temperature uniformity[23–25], and low emission power. Previous designs for thermal metasurfaces have been mainly focused on the infinite array of subwavelength scatters, which is beneficial for experimental realization with excellent optical properties[1,3,6], but leads to a slow response speed and a non-uniform temperature distribution.

In this work, we study the size effect of electrothermal metasurfaces by designing gold nanorod array (GNA) based metasurfaces with a nearly perfect narrowband emission, as shown in Fig. 1 (a). The resonance intensity from the finite GNA can be strongly converged to the value comparable to the infinite case. The resulting small footprint allows the metasurface thermal emitters to respond at a nanosecond timescale. More importantly, this strong resonance is to some extent insensitive to the aspect ratio of the nanorod array, which serves as a new degree of freedom to achieve the temperature uniformity of less than 1 K. The comprehensive analysis presented in this work thus provides a guideline for designing high-performance thermal infrared metasurfaces. A detailed benchmark discussion in Appendix shows that even state-of-the-art active thermal metasurfaces can still benefit from our finite-size and aspect-ratio analyses.

[*]E-mail: sshen1@cmu.edu

[†]The authors contribute equally to this work.

## 2 Modelling methods

The optical response of the metasurface is modelled in Ansys Lumerical finite-difference time-domain solutions. Then, electrothermal simulations are conducted in COMSOL Multiphysics. The GNA is positioned on the multilayer consisting of a Si substrate, a 300 nm thick layer of $SiO_2$ insulation layer, a 200 nm thick Al reflector, and a 100 nm thick dielectric spacer made of $Al_2O_3$, as shown in Fig. 1 (b). The Al reflector also functions as a heater, where a current is passed through to heat up the GNA.

In the optical simulations, the GNA and beneath substrate are illuminated by a plane wave source polarized in the x-direction, as indicated by the blue arrow in Fig. 1 (b). A reflective power monitor is placed above the source to capture the reflective energy and thus the resonance feature of the GNA.

The thermal simulation combines an electric current module, governed by the electric current equations, with a 3D heat conduction module, governed by heat diffusion equations. In the electric current module, a fixed current of $I_0$ = 250 mA passes through the Al reflector, and the calculated volumetric Joule heat generation is set to be the heat source in the heat conduction module. The bottom surface of the silicon substrate is fixed to have a constant temperature of $T_0$ = 293.15 K. The transient simulation starts with the overall initial temperature of $T_0$, and the temperature profiles are captured within the time from 0 to 2000 ns.

## 3 Results and discussion

### 3.1 Optical simulations

To analyze the finite size effect on the metasurface, we first simulate the reflective spectrum of an infinite GNA as a reference. Typically for an infinite GNA, the resonance frequency in the reflective spectra is determined by the localized surface plasmon resonance (LSPR) of the single gold nanorod resonator[26], and the resonance strength is governed by both the LSPR and the packing density (periodicity)[27]. Here, we design the GNA with the single nanorod sized as 700 nm by 50 nm by 50 nm in the x, y, and z directions and the periodicities set to be Px = Py = 800 nm. The simulated spectrum is plotted as the black line in Fig. 2(a) and features a first-order resonant peak centered around 1046 nm in the wavelength. Figure 2 (b) illustrates the $|E|^2$ profile at the resonance frequency, which indicates a longitude LSPR mode[26]. High-order mode at around 870 nm in Fig. 2 (a) corresponds to the weak periodic brightness near the center of the nanorod in Fig. 2 (b).

The effect of the finite array size on the GNA metasurface is then investigated with a different finite number of N × N arrays from N = 1 to N = 9, as shown in Fig. 2 (a). Although for simulations involving finite size arrays, it is inherently impossible to eliminate the absorption effect of the simulation boundaries (perfectly matched layer)[28], which will inevitably result in an underestimation of the reflectance. We thus, in all finite array size simulations, fix the monitor size (fully covered the 9 by 9 arrays) and the distance between the monitor and the metasurface. By doing so, we can mimic the real experimental condition where those arrays are measured by the same spectrometer. Hence those reflectance spectra are comparable, and the spectral emissivity of those finite size metasurfaces can be inferred based on Kirchhoff's law.

From Fig. 2 (a), we can find that the finite array preserves the narrowband feature of the infinite one and the first-order resonance starts to saturate when N = 7, which corresponds to a metasurface with an area of only about 2.2 μm². This enables the device to have a thermal response down to nanoseconds as discussed later in electrothermal simulations. Compared to the case of infinite array size, a red shift of resonance can be clearly observed. It depends on the packing density or periodicity of the nanorod array, as described by the tight-binding model[27]. With an increasing number of nanorods, the dipole moment can be effectively elongated, causing a longer resonance wavelength.

The influence of the GNA arrangement to the reflectance spectra is also studied. To achieve this, we keep the total number of unit cells around 49, which corresponds to the 7 by 7 array marking the resonance saturation, but vary the aspect ratios Nx/Ny of the array (i.e., number of unit cells along the axial direction (x-direction) divided by the number along the other (y-direction)). For arrays with different aspect ratios, the contrast ratio Cr of the reflectance spectrum, defined as:

$$Cr = [1-\min(R)]/[1-\text{mean}(R)] \quad (1)$$

is evaluated. In Eq. (1), min(R) represents the minimum value in the reflectance spectrum and mean(R) represents the spectral average of the reflectance spectrum. Hence, the contrast ratio physically represents that to which degree the resonance feature stands out of the baseline of the spectrum. Therefore, a large Cr value is targeted.

From Fig. 3 (a), for Nx/Ny > 1, the resonance can maintain its intensity even with only a small number of arrays in the perpendicular direction (y-axis) due to the highly polarized nature of the GNA resonance. On the other hand, the arrays with an aspect ratio Nx/Ny < 1 show a significantly decreased contrast ratio. A detailed comparison can be found in Fig. 3 (b), where Nx/Ny = 16/3 almost keeps the same resonance intensity as that of the unbiased case Nx/Ny = 7/7 while the peak height of Nx/Ny = 3/16 is reduced by almost a half and broadened. The relaxation of the requirement for the number of arrays in the direction perpendicular to polarization can be found later to significantly improve the temperature uniformity of the GNA when considering the electrothermal active emission. It is thus a valuable factor in design, especially at the nanoscale where the controlling parameters are limited.

*3.2 Thermal simulations*

The GNA is then electrically heated, and its emission is analyzed in both transient and steady states. We first simulate the response time (defined as 90% of the rise time responding to the on-switch current injection) for the device under different array sizes and aspect ratios. With the increased array size, the response time increases dramatically, as shown in Fig. 4, since a larger size induces a larger thermal mass which significantly reduces the cutoff modulation frequency. This shows the importance of finding the minimal size of the finite metasurface showing enough resonance contrast. For our device, with a saturation size of N = 7, we can achieve a response time of about 600 ns, which outperforms the majority of the current tunable metasurface designs (see Appendix).

Under the similar array size around N = 7, containing 49 unit cells, we further study the influence of the aspect ratio. Even under an extremely large aspect ratio change (Nx/Ny = 4/12 ~ 12/4), we can keep the device response time down to nanoseconds, with a fluctuation of only around 100 ns (red-dashed box).

Combined with our previous discussion that a large Nx/Ny has less influence on the resonance contrast of the x-polarized metasurface, the aspect ratio can thus hold the metasurface a stable performance both optically with a narrowband, nearly perfect emission and electrothermally with an ultrafast response.

Therefore, under a certain window, the aspect ratio can be set freely to benefit other aspects of our device performance, among which the temperature uniformity of the metasurface is fundamentally important. A reliable thermal emission typically favors a stable and uniform temperature, which also simplifies the design processes. Even when pursuing some advanced nonequilibrium emission[29], controlling temperature uniformly at a small local area is a prerequisite for most cases.

The temperature uniformity of the GNA metasurface is then illustrated by the 2D normalized isothermal contours, as shown in Fig. 5 (a), together with the geometry of the arrays and electrodes. The GNA with a larger Nx/Ny is found to have much sparser isothermal contours, indicating a smaller temperature gradient. To specify the details, we also define the temperature difference (Tmax – Tmin) inside the array area enclosed by black-dashed boxes in Fig. 5 (a). The temperature difference for both axes, as shown in Fig. 5 (b), decrease with increasing Nx/Ny because of the thermal conduction from the electrode, which substantially modifies the temperature profile of the metasurface. Another indication of the impact from the electrode is that the decrease of temperature difference along the y-axis is faster than that of the x-axis due to the more effective heat conduction through the electrode metals. With Nx/Ny increasing up to about 3, the temperature difference can even reach sub-Kelvin along the y-axis while keeping a high resonance contrast ratio of 1.8 and a fast response time of 720 ns. It should be noted that here we only consider a common trapezoidal electrode, whose material and shape can be further optimized together with the GNA aspect ratio to enhance the temperature uniformity. Finally, the degree of freedom in aspect ratio tuning can also improve the power injection into the metasurface by the impedance matching based on the maximum power transfer theorem.

## 4 Conclusions

In summary, we design a thermal metasurface based on a finite-size gold nanorod array, which, with only several unit cells, can achieve a nearly perfect narrowband emission close to that of the infinite case. Because of its small volume, the metasurface can achieve a fast response down to a nanosecond level. Moreover, the aspect ratio of this finite array can be tuned to greatly enhance the temperature uniformity to a sub-Kelvin temperature difference while maintaining the high resonance contrast ratio and the nanosecond response time. The analysis methodology can be extended to polarization-independent metasurfaces, such as the crossbar array that shows better emission efficiency in sensing applications[16,30–33]. Furthermore, our analysis can also be extended to handle inhomogeneous temperature cases under the framework of a local Kirchhoff's law[34]. With these extensions, the analyses of the size and aspect ratio of the finite-size metasurfaces thus pave the way to optimize the electrothermal infrared devices.

**Appendix:**

The modulation of thermal radiation by a metasurface can be classified into two mechanisms: temperature and emissivity modulation, both of which can benefit from our finite-size and aspect-ratio analyses. To the best of our knowledge, we have summarized state-of-the-art active thermal metasurface in the table below.

| Year | Mechanism | Medium | Wavelength | Response Time | Reference |
|------|-----------|--------|------------|---------------|-----------|
| 2013 | Emissivity | MEMs | 6.2 μm | 30 kHz (33.3 μs) | [35] |
| 2017 | Emissivity | MEMs | 8.9 μm | 100 kHz (9.1 μs) | [36] |
| 2014 | Emissivity | GaAs/AlGaAs | 9.17 μm | 600 kHz (1.7 μs) | [37] |
| 2018 | Emissivity | InAs | 7.3 μm | 4.8 MHz (208.3 ns) | [22] |
| 2019 | Emissivity | GaN/AlGaN | 4 μm | 50 kHz (20.0 μs) | [38] |
| 2013 | Emissivity | Graphene | 7.8 μm | 40 MHz (25.0 ns) | [39] |
| 2014 | Emissivity | Graphene | 6.9 μm | 20 GHz (0.05 ns) | [40] |
| 2016 | Emissivity | Graphene | 8 μm | 2.6 GHz (0.38 ns) | [41] |
| 2018 | Emissivity | Graphene | 8.5 μm | 7.2 GHz (0.14 ns) | [42] |
| 2021 | Emissivity | GSST | 1.43 μm | 500 ms | [23] |
| 2021 | Emissivity | GST-326 | 755 nm | 21 μs | [24] |
| 2022 | Emissivity | GST-225 | 1.64 μm | 200 μs | [25] |
| 2019 | Temperature | Hot Electrons | 1.59 μm | 350 ps | [43] |
| 2015 | Temperature | Heat Diffusion | 4.26 μm, 3.95 μm | 20 Hz (50 ms) | [44] |
| 2018 | Temperature | Heat Diffusion | 4.2 μm, 7 μm | 100 kHz (10 μs) | [45,46] |
| 2021 | Temperature | Heat Diffusion | 5.1 μm | 20 MHz (50 ns) | [34] |

The emissivity modulation by MEMS[35,36], semiconductor carrier injection[22,37,38], and graphene gating[39–42] are pure electrical processes whose response times are limited by the electrical RC time constant of the device. This is also for the temperature modulation by hot electrons[43], which is the non-resonant emission when electrons and phonons are in non-equilibrium. Therefore, their response times are generally much shorter than our electrothermal case. Nevertheless, for our optical part of conclusions, the fast resonance convergence of finite-size array and the resonance insensitivity to aspect-ratio, are still applicable to these devices to reduce the response time by further eliminating the electrical parasitic capacitance.

The response time for the emissivity modulation by phase-change materials[23–25] is limited by the set pulse which needs to be long enough for the GeTe-based materials fully crystallized. This mechanism is particularly suitable for our guidelines regarding both the optical part and the thermal part. The response time can be further decreased, and the temperature uniformity improves as the array size decreases.

The temperature modulation by heat diffusion[34,44–46] is our ideally targeted case. Our work focuses on the metasurface design to handle the tradeoffs between the optical and thermal performances, such as the tradeoff between the infinite-array requirement in optical responses and the high speed in thermal responses. However, the response time is a systematic result not only from the metasurface layer but also from the whole device including the electrode design for heat generation and the substrate for heat dissipation. Thus, it is understandable that the work [34], after careful optimization of all these factors, has achieved a response time of 20ns.

Although the temperature uniformity is not comprehensively measured from the work above, it keeps fundamentally important since most metasurface-based thermal radiation control implicitly assumes the validity of Kirchhoff's law[22–25,35–46]. Otherwise, a generalization of Kirchhoff's law[34] or some inhomogeneous direct emission computations[29] must be included, increasing the difficulties in design. The temperature uniformity is particularly critical in the phase-change-based modulation, which requires the heat pulse not only appropriate in the time domain but also active in optically functional areas in the space domain.

**Acknowledgements** This work was primarily supported by the Defense Threat Reduction Agency (Grant No. HDTRA1-19-1-0028). This work was also funded partially by the National Science Foundation (Grant No. CBET-1931964). X. L. and Z. L. identified the problem; Z. L. and X. L. conducted the optical simulations; Z. W. and X. L. conducted the thermal simulations; X. L. and Z. L. prepared the manuscript with the input from Z. W. and H. S. Y.; S. S. supervised the research. All authors have given approval to the final version of the manuscript. The authors declare no competing financial interest.

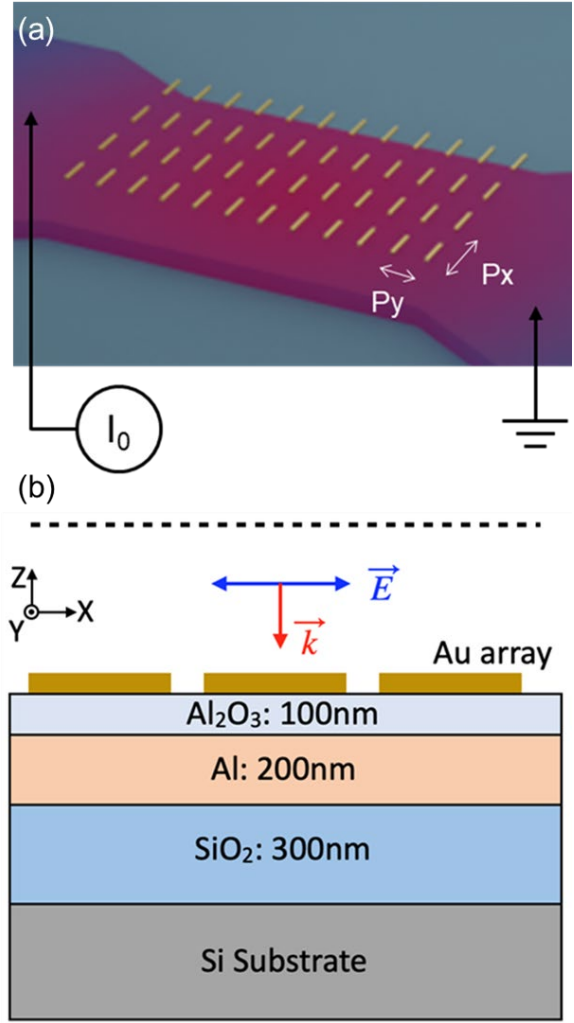

**Fig. 1** (a) Schematic of a metasurface based on gold nanorod arrays. The $Al_2O_3$ spacer is hidden to reveal the beneath Al reflector, which also functions as the heater (Joule heating is applied via a fixed current $I_0$). Px and Py represent the periodicities in the x-direction and the y-direction, respectively. (b) x-z view of the simulated device. The dimensions of each gold nanorod are 700 nm by 50 nm by 50 nm in the x, y, and z directions. The size and the location of the reflective power monitor (dashed line) is fixed among simulations.

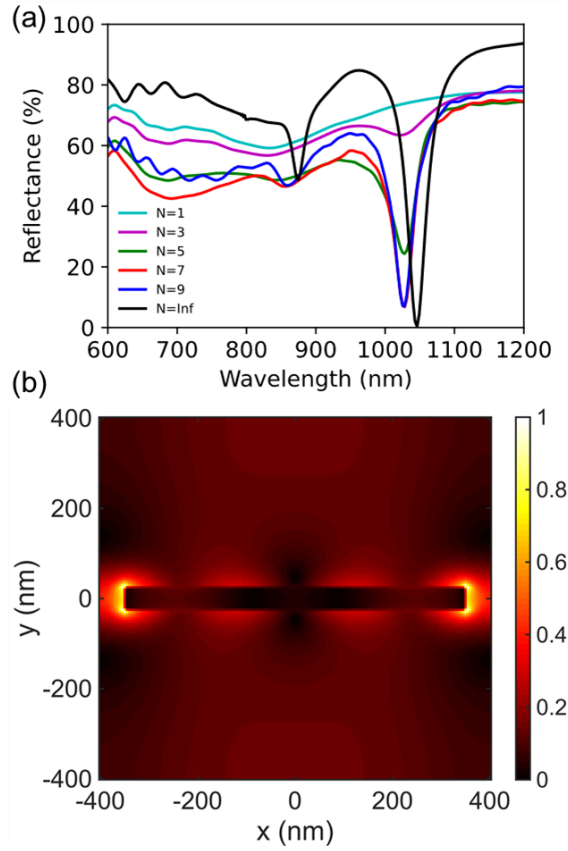

**Fig. 2** (a) Simulated reflectance spectra for different array sizes. (b) Simulated $|E|^2$ profile of a single unit cell in the infinite array at the resonance wavelength of 1046 nm.

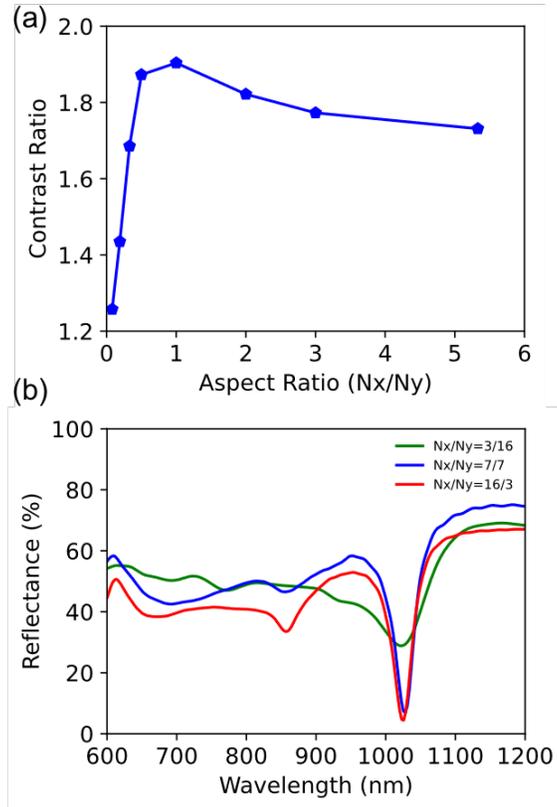

**Fig. 3** (a) Contrast ratios $C_r$ of finite GNAs with different aspect ratios. (b) Simulated reflectance spectra for aspect ratios of Nx/Ny = 16/3, 7/7 and 3/16, respectively.

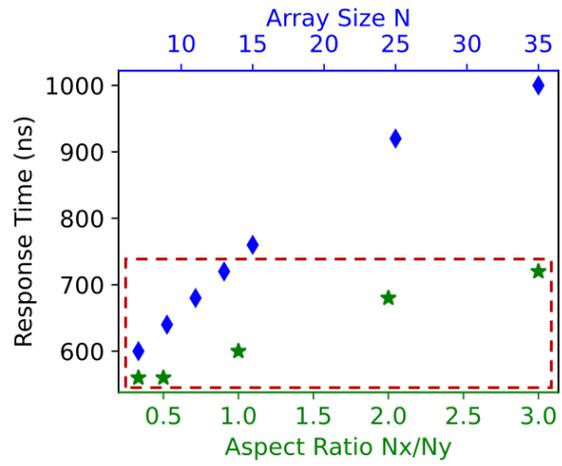

**Fig. 4** Response time for arrays in different sizes N (blue diamonds) and aspect ratios Nx/Ny (green stars).

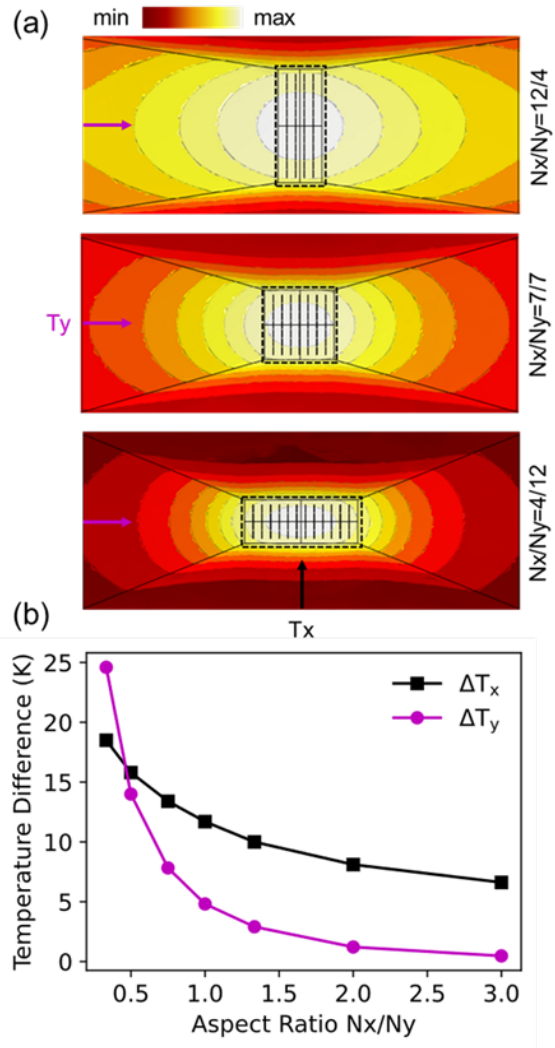

**Fig. 5** (a) From top to bottom: normalized isothermal contours for arrays in aspect ratio Nx/Ny = 12/4, 7/7, and 4/12, respectively. (b) Temperature difference along x-axis $\Delta T_x$ and along y-axis $\Delta T_y$ for arrays in different aspect ratios.